\input epsf
\def\tightitemize{\begin{itemize}\setlength{\parskip}{0.5 \parsep}
                  \setlength{\itemsep}{0 pt}}

\def\endcvitemize{\end{itemize} \end{itemize} \end{itemize}}
\def\selectedbib{\section[Selected Bibliography]
{Selected Bibliography}
\bgroup\parindent=0pt\parskip=\itemsep
\def\refpar{\par\hangindent=3em\hangafter=1}}
\def\endselectedbib{\refpar\egroup}

\def\agt{>\kern-1.1em \lower1.1ex\hbox{$\sim$}\kern.3em}
\def\alt{<\kern-1.1em \lower1.1ex\hbox{$\sim$}\kern.3em}
\def\kms{${\rm km~s^{-1}}\;$}
\def\3he{$^3{\rm He}$}

\def\gs{\mathrel{\raise0.35ex\hbox{$\scriptstyle >$}\kern-0.6em 
\lower0.40ex\hbox{{$\scriptstyle \sim$}}}}
\def\ls{\mathrel{\raise0.35ex\hbox{$\scriptstyle <$}\kern-0.6em 
\lower0.40ex\hbox{{$\scriptstyle \sim$}}}}
\newfont{\rmsmall}{cmr12 scaled 900}

\def\chigh{${ {}^{\rm 3}P_{\rm 2}\rightarrow{}^{\rm 3}P_{\rm 1}}$}
\def\jeqf{$J=4\rightarrow3$}
\def\jeqs{$J=7\rightarrow6$}
\def\jeqo{$J=1\rightarrow0$}
\def\jeqt{$J=2\rightarrow1$}
\def\litl{\rm\scriptscriptstyle}

\def\carbon#1{\ifmmode{{^{#1}{\rm C}}}\else{{$^{#1}{\rm C}$}}\fi}

\def\nnh{\ifmmode{{{\rm N}_{\litl H}}}\else{{${\rm N}_{\litl H}$}}\fi}

\def\nc{\ifmmode{{{\rm N}_{\litl C}}}\else{{${\rm N}_{\litl C}$}}\fi}
\def\nco{\ifmmode{{{\rm N}_{\litl CO}}}\else{{${\rm N}_{\litl CO}$}}\fi}

\def\ci{C~{\rmsmall I}}

\def\nii{N~{\rmsmall II}}

\def\car#1 {\alwaysmath{{}^{#1}{\rm C}}}

\documentstyle[aaspp4, 12pt, hacks, references]{article}
\renewcommand{\thesubsection}{\arabic{subsection}}

\begin{document}
\title{\Large Terahertz Initiatives at the \\
Antarctic Submillimeter Telescope and Remote Observatory (AST/RO)}
\vskip 16pt
\noindent
\author{$\underline{\rm Antony~A.~Stark}$, Adair P. Lane, Christopher L. Martin (SAO),\\
Richard A. Chamberlin, Jacob Kooi (Caltech), \\
and Christopher K. Walker (U. Arizona)
}
\vskip 20pt
\begin{abstract}{
\noindent
The Antarctic Submillimeter Telescope and Remote Observatory (AST/RO)
is a 1.7-meter diameter offset Gregorian instrument located at the NSF 
Amundsen-Scott South Pole Station.  This site is exceptionally dry and 
cold, providing opportunities for Terahertz observations from the ground.  
Preliminary analysis of recent site testing results shows that the zenith 
transparency of the 1.5~THz atmospheric window at South Pole frequently 
exceeds 10\% during the Austral winter.  Routine observations at 
810~GHz have been conducted over the past two years, resulting in 
large-scale maps of the Galactic Center region and measurements of the
$^{13}$C line in molecular clouds.  During the next two years, the observatory 
plans to support two Terahertz instruments: \hfill

1)\quad TREND ({\it Terahertz Receiver with Niobium Nitride Device}---K. S. \mbox{Yngvesson}, University of
Massachusetts,
P. I.), and \hfill

2)\quad SPIFI ({\it South Pole Imaging Fabry-Perot Interferometer}---G. J. Stacey, Cornell University, P. I.).
\hfill

\noindent
AST/RO could be used in future as an observational test bed for additional
prototype Terahertz instruments.  Observing time on AST/RO is available on
a proposal basis (see {\tt http://cfa-www.harvard.edu/$\sim$adair/AST\_RO}).
}
\end{abstract}

\textwidth=6.5in

\section{The AST/RO Instrument}
\label{subsec:description}

The Antarctic Submillimeter Telescope and Remote Observatory (AST/RO) 
is an instrument routinely used for the measurement of
submillimeter-wave spectral lines over regions 
several square degrees in size
toward the Milky Way and Magellanic Clouds.
AST/RO is a 1.7m ~diameter offset Gregorian telescope, with
optics designed for 
wavelengths between 200 \micron \ and 3 mm. 
All of the optics in AST/RO are offset for high beam efficiency and avoidance of inadvertent reflections and resonances.
The design of AST/RO is described in \cite{stark97a}. 
AST/RO site testing, logistics, capabilities,
and observing techniques are described in \cite{stark01}.

Currently, there are five heterodyne receivers mounted on an
optical table suspended from the telescope structure
in a spacious
($5 \mathrm m \times 5 \mathrm m \times 3 \mathrm m$), warm Coud\'{e}
room:
\begin{enumerate}
\item {a 230 GHz SIS receiver, 85 K double-sideband (DSB) noise temperature (\cite{kooi92});}
\item {a 450--495 GHz SIS quasi-optical receiver, 
165--250 K DSB (\cite{engargiola94,zmuidzinas92});}
\item {a 450--495 GHz SIS waveguide
receiver, 200--400 K DSB (\cite{walker92,kooi95}), which can be
used simultaneously with} 
\item  {a 800-820 GHz fixed-tuned
SIS waveguide mixer receiver, 950--1500 K DSB (\cite{honingh97});}
\item {an array of four 800-820 GHz fixed-tuned
SIS waveguide mixer receivers, 850--1500 K DSB
(the PoleSTAR array, see {\tt http://soral.as.arizona.edu/pole-star} 
and
\cite{groppi00}).}
\end{enumerate}
Spectral lines observed with AST/RO include:
CO \jeqt ,
CO \jeqf ,
CO \jeqs ,
HDO $J = 1_{0,1} \rightarrow 0_{0,0}$,
[\ci \,] $^3P_1 \rightarrow {}^3P_0$, 
[\ci \,] $^3P_2 \rightarrow {}^3P_1$, and
[${}^{13}$\ci \,]  $^3P_2 \rightarrow {}^3P_1$.
A proposal is currently pending to the Smithsonian Institution
to purchase a local oscillator to cover 650--700 GHz,
a frequency range which 
includes the $^{13}$CO $J = 6 \rightarrow 5$ line.
There are four currently available
acousto-optical spectrometers (AOS), all designed and built at
the University of Cologne (\cite{Schieder89}):
two low-resolution spectrometers
with a bandwidth of 1 GHz (bandpass 1.6--2.6 GHz); 
an array AOS having four low
resolution spectrometer channels with a bandwidth of 1 GHz (bandpass
1.6--2.6 GHz)
for the PoleSTAR array;
and 
one high-resolution AOS with 60 MHz bandwidth
(bandpass 60--120 MHz). 

AST/RO has been open to proposals from the general astronomical community since 1997.
AST/RO research is
a three part effort, where
approximately equal time is given to each of these 
initiatives:
\begin{enumerate}
\item large-scale surveys of regions of general interest:
the Galactic Center and the  Magellanic Clouds;
\item support of observations of special interest, through 
observing proposals solicited from the worldwide astronomical community;
\item support of technology development, by making the telescope
available for installation and trial of novel detectors, especially
detectors at Terahertz frequencies.
\end{enumerate}

\section{Site Testing}

\begin{figure}[tb!]
\begin{center}
\begin{minipage}{3.7in}
\vskip 10 pt
\epsfxsize=3.5in
\ \ \epsfbox{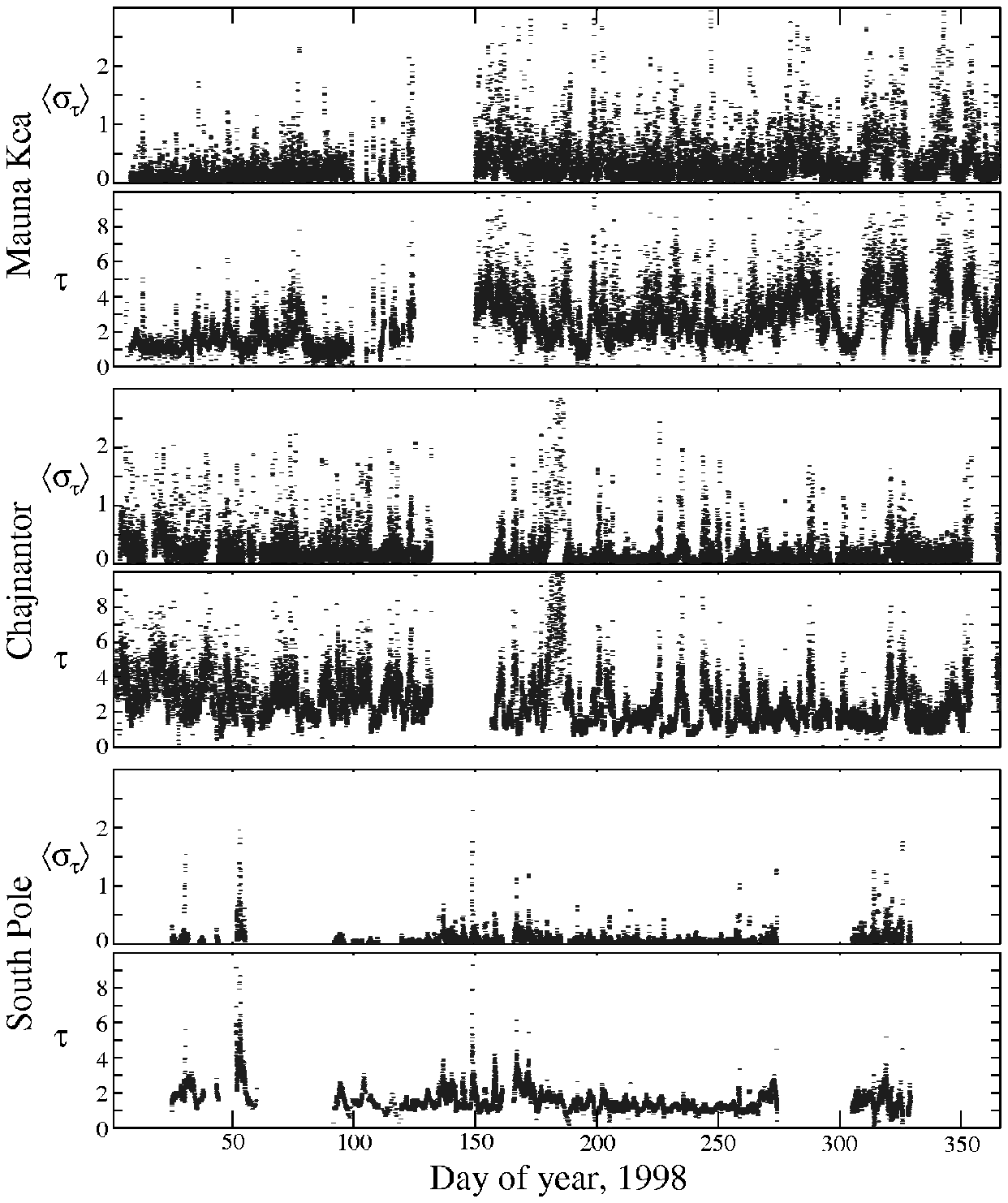}
\vskip 20 pt
\end{minipage}
\begin{minipage}{2.4in}
\caption[Sky Noise and Opacity Measurements at 350 \micron \ from
Three Sites]
{{\bf Sky Noise and Opacity Measurements at 350 \micron \ from Three
Sites.\ \ }
These plots show data from identical NRAO-CMU 350 \micron \ broadband
tippers (Radford \& Peterson, unpublished data)
located at Mauna Kea, Hawaii; the ALMA site at Chajnantor, Chile; and South Pole
during 1998.  The upper
plot of each pair shows $\langle \sigma_\tau \rangle$, the rms
deviation
in the opacity $\tau$ during a one-hour period---a measure
of sky noise on large scales; the lower plot of each
pair shows $\tau$, the broadband 350 \micron \ opacity.
The first 100 days of 1998 on Mauna Kea were exceptionally good for
that
site.  During the best weather at the Pole,
$\langle \sigma_\tau \rangle$ was dominated by detector noise rather
than
sky noise.
\label{fig:threesites}
\vfill}
\vskip 10 pt
\end{minipage}
\end{center}
\end{figure}

\begin{figure}[t!]
\begin{center}
\leavevmode
\epsfxsize=5.5in
\epsfbox{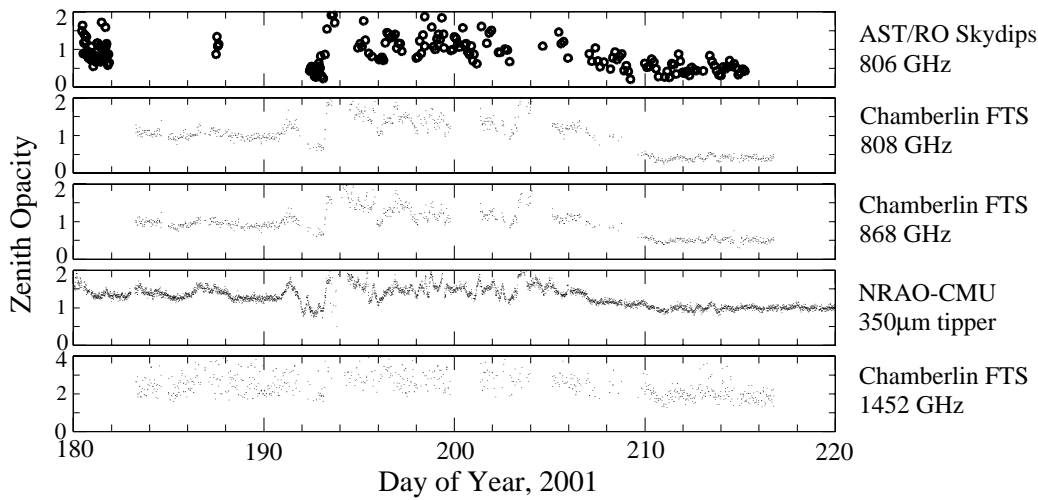} 
\end{center}
\begin{center}
\parbox[t]{5.98in}{
\caption[Simultaneous Opacity Measurements from Three Instruments at the South Pole] 
{\small {\bf Simultaneous Opacity Measurements from Three Instruments 
at the South Pole.} These plots show data from 
AST/RO skydips, the NRAO-CMU 350 $\mu \rm m$ broadband
tipper (Radford \& Peterson, unpublished data), 
and a  submillimeter-wave Fourier Transform
Spectrometer built by Richard Chamberlin
in July and early August of 2001.
The AST/RO data are from skydips taken for calibration
purposes during observations; they agree well with the
FTS measurements at 808 GHz.  The FTS measurements at 868 GHz
are shown for comparison with the NRAO-CMU broadband
measurements, which are centered at that frequency.
Note that the NRAO-CMU tipper values are monotonically
related to the FTS measurements, but show an offset
and compression of scale.  The bottom plot
shows a preliminary reduction of 
FTS measurements at 1.452 THz, and indicates $\tau < 2$
for a significant fraction of the time.  Usually August
and September are the best months at the Pole; these
observations unfortunately had to
be stopped early because of insufficient
liquid helium supplies.
\label{fig:threeinstruments}
}
}
\end{center}
\end{figure}

\begin{figure}[tb!]
\begin{center}
\begin{minipage}{3.7in}
\vskip 10 pt
\epsfxsize=3.2in
\ \ \epsfbox{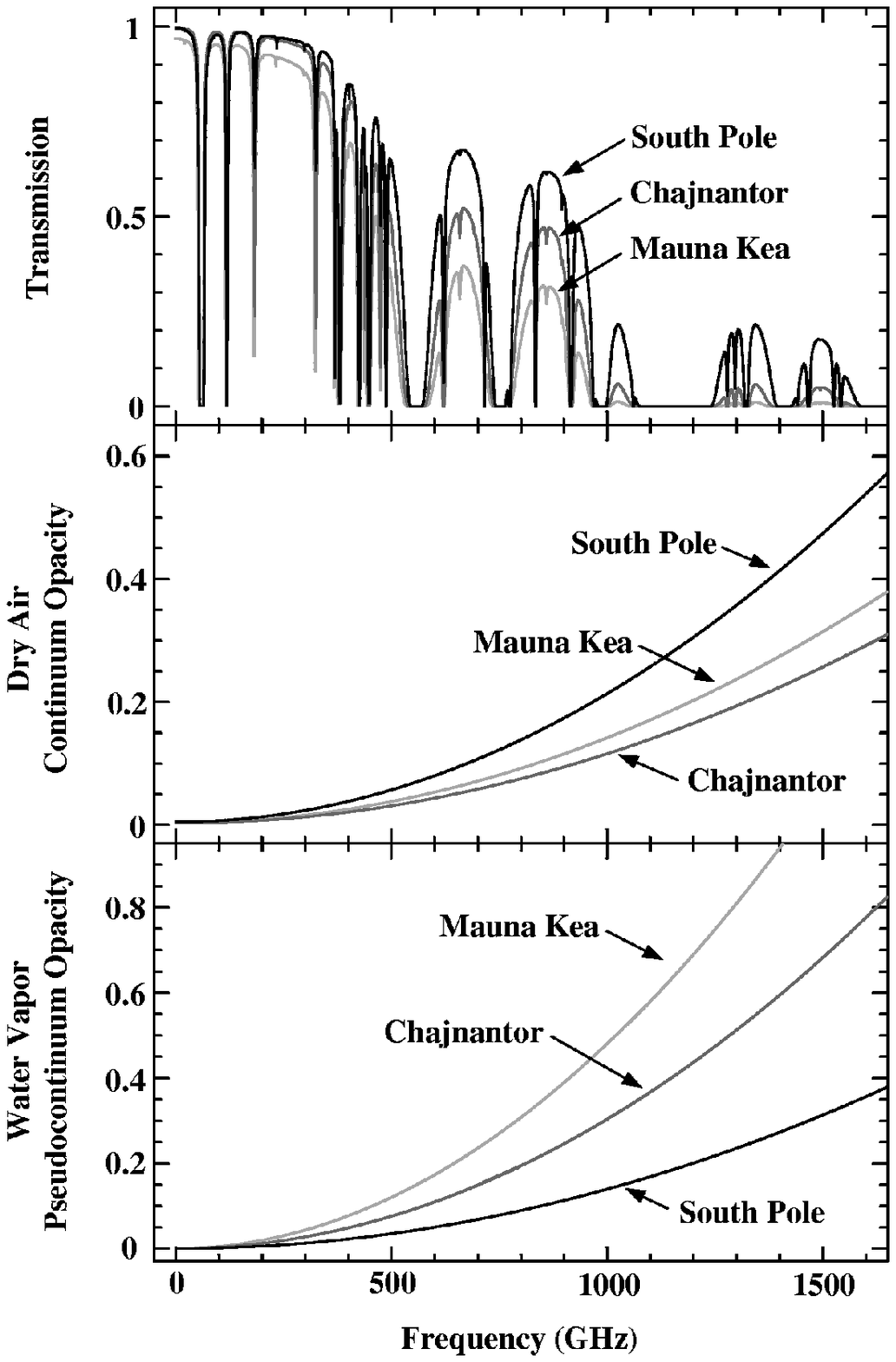}
\vskip 20 pt
\end{minipage}
\begin{minipage}{2.4in}
{
\caption[Calculated atmospheric transmittance at three sites.\ \ ]
{\small {\bf Calculated atmospheric transmittance at three sites.\ \ }
The upper plot is atmospheric transmittance at zenith
calculated by J. R. Pardo using the ATM model (\cite{pardo01b}).
The model uses PWV values of 0.2 mm for South Pole, 0.6 mm
for Chajnantor and 0.9 mm for Mauna Kea, corresponding to
the $25^{\mathrm{th}}$ percentile winter values at each site.
Note that at low frequencies, the Chajnantor curve converges with the
South Pole curve, an indication that 225 GHz opacity is not a simple
predictor of submillimeter wave opacity.
The middle and lower plots show calculated values of dry air
continuum opacity and water vapor pseudocontinuum opacity for
the three sites.  Note that unlike the other sites, the opacity
at South Pole is
dominated by dry air rather than water vapor.
\label{fig:pardo}
\vfill}
\vskip 10 pt
}
\end{minipage}
\end{center}
\end{figure}

The South Pole is an excellent millimeter- and submillimeter-wave site 
(\cite{lane98,chamberlin94,chamberlin97,chamberlin02}).
It is unique among observatory sites for unusually low
wind speeds, absence of rain, and the consistent clarity of the
submillimeter sky.  
\cite{schwerdtfeger}
has comprehensively reviewed the climate of the Antarctic Plateau
and the records of the
South Pole meteorology office.
Chamberlin (2001)
has analyzed weather data to determine the precipitable
water vapor (PWV) and finds median wintertime 
PWV values of 0.3 mm over a 37-year period, with little annual
variation.
{\em PWV values
at South Pole are small, stable, and well-understood.}

Sub\-millimeter-wave atmospheric
opacity at South Pole has been measured using skydip techniques.
We made
over 1100 skydip observations at 492~GHz (609 \micron ) with AST/RO
during the 1995 observing season (\cite{chamberlin97}).
Even though this frequency is near a strong oxygen line, the opacity
was
below 0.70 half of the time during the Austral winter and reached
values as
low as 0.34, better than ever measured at any ground-based site. The
stability was also remarkably good: the opacity remained below 1.0 for
weeks
at a time.
From early 1998, the 350\micron \, band has been continuously
monitored
at Mauna Kea, Chajnantor, and South Pole by identical tipper
instruments
developed by S. Radford of NRAO and J. Peterson of Carnegie-Mellon U.
and
the Center for Astrophysical Research in Antarctica (CARA).  
Results from Mauna Kea and Chajnantor are compared with South Pole in
Figure~\ref{fig:threesites}.
{\em The 350\micron \ opacity at the South Pole is consistently better
than at Mauna Kea or Chajnantor.}

A new Fourier Transform Spectrometer developed by R. Chamberlin and
collaborators was operational at the Pole during some of the 
winter of 2001.  This instrument measures a broadband spectrum
covering 300 GHz to 2 THz as a function of airmass several times
each hour.  Some of these data are shown in Figure 2.  The
zenith transparency at 1.452 THz (near an important [\nii \,]
line) exceeded 10\% for almost the entire first week of August 2001.
{\em The observed relation between the NRAO-CMU tipper and the
1.452 THz measurements indicates that it will be possible to
observe the $\lambda 205\, \mu {\rm m}$} {\rm [\nii \,] } {\em line about 30 days each year.}

The South Pole 25\% winter PWV levels
have been used to compute values of atmospheric
transmittance
as a function of wavelength which are plotted in
Figure~\ref{fig:pardo}. For comparison, the transmittances for
25\% winter conditions at Chajnantor and Mauna Kea are also shown.

{\em Sky noise} is caused by fluctuations
in total power or phase of a detector caused by
variations in atmospheric emissivity and path length on timescales of
order one second.
Sky noise causes systematic errors in the measurement of astronomical
sources.  
\cite{lay99} show analytically how sky noise causes observational
techniques
to fail: fluctuations in a component of the data due to sky noise
integrates
down more slowly than $t^{-1/2}$
and will come to dominate the error during long
observations. 
Sky noise at South Pole is considerably smaller than at other sites,
even comparing conditions of
the same opacity.  
The PWV at South Pole is
often so low that the opacity is dominated by the {\em dry air}
component (\cite{chamberlin95,chamberlin01}, cf. Figure
\ref{fig:pardo});
the dry air emissivity
and phase error do not vary as strongly or rapidly as
the emissivity and phase error due to water vapor.

\section{810 GHz Observations at AST/RO}

AST/RO has detected the isotopic [$^{13}$\ci \ ] \chigh~ fine-structure 
transition in three galactic regions:
G 333.0-0.4, NGC 6334 A, and G 351.6-1.3.
This is only the second time that this line 
have been successfully observed, the previous detection being 
a single spectrum obtained with the Caltech Submillimeter Observatory
toward the Orion Bar (\cite{keene98}). 
The [$^{13}$C I] line was
observed simultaneously with the CO \jeqs ~ line emission at 806 GHz
(\cite{tieftrunk01}). 

Essentially all of the NGC 6334 Giant Molecular Cloud was mapped  
in 492 and 810 GHz [\ci ] and the CO \jeqs ~ and \jeqf ~
spectral lines.  
The data show that high
excitation temperatures exist throughout most of the cloud volume.
Detailed modeling is in progress to account for the observed line
intensities and ratios (Yan et al. in preparation).

An up-to-date bibliography of AST/RO publications
can be found at
the AST/RO website {\tt http://cfa-www.harvard.edu/$\sim$adair/AST\_RO}.

\begin{figure}[tb!]
\begin{center}
\begin{minipage}{3.5in}
\vskip 10 pt
\epsfxsize=3.3in
\ \ \epsfbox{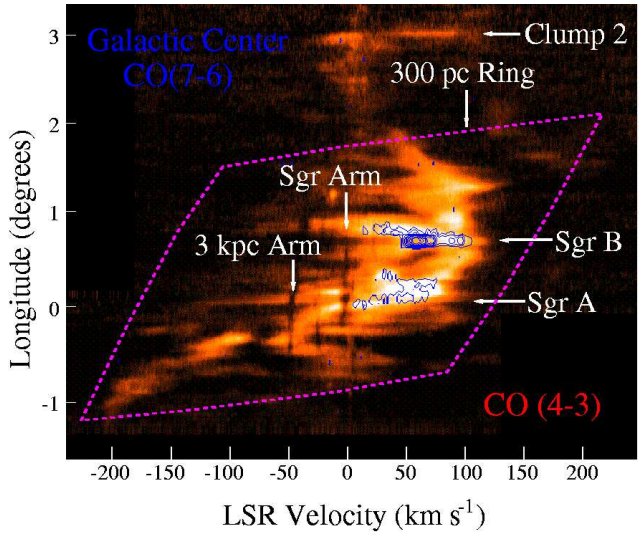}
\vskip 20 pt
\end{minipage}
\begin{minipage}{2.3in}
\caption[AST/RO observations of the Galactic Center Region]
{\small {\bf AST/RO observations of the Galactic Center Region}
(from \cite{kim00}).
The CO \jeqf~ (pseudo-color) and \jeqs~ (blue contour) lines
observed in an
$l - v$ strip, sampled every $1'$, at $b = 0$.
These data have been used in conjunction with
CO  and $^{13}$CO \jeqo~ data to determine the
the density and temperature in the features shown here.
\label{fig:GCsung}
\vfill}
\vskip 10 pt
\end{minipage}
\end{center}
\end{figure}

\suppressfloats

\section{Terahertz Initiative}
\label{sec:terahertz}

\noindent
Two new short wavelength instruments are in development for use on
AST/RO:
\begin{itemize}
\item{
Dr. G. Stacy and collaborators have developed the South Pole Imaging
Fabry-Perot Interferometer (SPIFI, \cite{swain98}), 
a 25-element bolometer array
preceded by a tunable Fabry-Perot filter.  This instrument
was successfully used on the JCMT in May 1999 and April 2001 and is being
modified with new instrumentation, cryogenics, and detectors
for South Pole use. SPIFI is frequency agile and can observe many
beams at once, but has limited frequency resolution ($\sim 100$ \kms )
and scans a single filter to build up a spectrum.
}
\item{
Dr. S. Yngvesson and collaborators (\cite{gerecht99,yngvesson01}) 
are developing a 1.5 THz heterodyne 
receiver, the Terahertz Receiver with Niobium Nitride Device (TREND).
TREND has only a single pixel and is not frequency agile.  Its HEB
device requires high local oscillator power levels;  we will use a
laser local oscillator source which requires that the gas be changed
in order to change frequencies.  The frequency resolution of
TREND is high ($\sim$ 2 MHz), limited by the stability of the laser.
}
\end{itemize}
In addition, Dr. D. Prober and collaborators are developing a low-noise,
low-power 1.5 THz heterodyne receiver based on aluminum and
tantalum HEB technology which may be tested on AST/RO.

\begin{figure}[tb!]
\begin{center}
\leavevmode
\epsfxsize=3.9in
\epsfbox{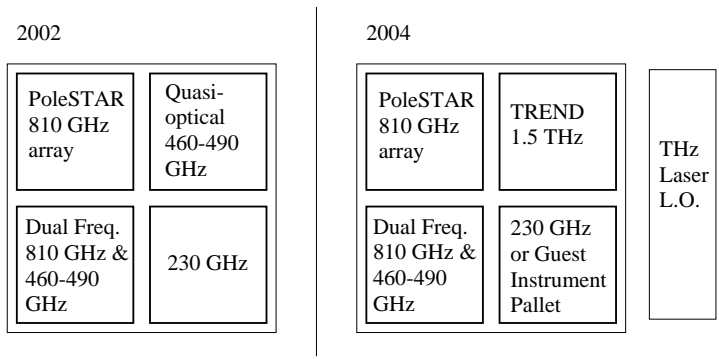} \\
\parbox[t]{6.00in}{
\caption[AST/RO Receiver Room Configurations]
{ \small {\bf AST/RO Receiver Room Configurations.\ } 
On the left is the current configuration of
pallets in the AST/RO receiver room, looking downwards to
the Coud\'{e} focus.
At right is a proposed configuration circa 2004,
with the TREND 1.4 THz HEB mixer and its
laser local oscillator installed.
A ``Guest Instrument Pallet"
would permit testing of high-frequency prototypes.
\label{fig:receiver} }
}
\end{center}
\end{figure}

Deployment of these technologies on a ground-based telescope 
is a path of technological development that has exciting
prospects.  On AST/RO, the $\sim 35^{\prime\prime}$ beam size
and high spectral resolution ($\sim 0.4$~ \kms) of Terahertz
receivers will allow the study of galactic star-forming
regions and large-scale studies of nearby galaxies.
In future, these detectors could be used on the South Pole
Submillimeter Telescope (SPST), an 8-meter telescope
(see \cite{nrc01}), which would have a beamsize of $\sim 7''$.

\nopagebreak
\section{Conclusion}
We hope to begin ground-based Terahertz observations with 
AST/RO in 2003.  Routine observations have been
carried out in the 350 \micron \,  window over the past two years,
and our experience has been that such observations are
possible more than 100 days each year.  Site testing
with a new Fourier Transform Spectrometer, combined with
long-term measurements from the NRAO-CMU tipper, indicates that
observations in the 200 \micron \,  window should be possible
about 30 days each year.  Two Terahertz detector systems,
SPIFI and TREND, are scheduled for installation in the next
two years.  We expect that after initial tests and observations
these instruments will become available for astronomical use
on a proposal basis.  AST/RO is also open to proposals
for tests of other prototype Terahertz instruments
in the coming years.

We thank Simon Radford of NRAO and Jeff Peterson of CMU for the data
shown in Figure \ref{fig:threesites}.
We thank Juan R. Pardo of Caltech for discussions on
atmospheric modeling and for carrying out
the calculations shown in Figure \ref{fig:pardo}.
The AST/RO group is grateful for the logistical support of the 
National Science Foundation, Antarctic Support Associates, 
Raytheon Polar Services Company,
and CARA during our polar expeditions.
This work was supported in part by  United States
National Science Foundation grant DPP88-18384, and by
the Center for Astrophysical Research in Antarctica and the NSF
under Cooperative Agreement OPP89-20223.


\begin{thebibliography}

\bibitem[{Chamberlin 2001}]{chamberlin01}
Chamberlin, R.~A. 2001, J. Geophys. Res., 106 (D17), 20101

\bibitem[{Chamberlin 2002}]{chamberlin02}
Chamberlin, R.~A. 2002, in ASP Conference Series, Vol. 266,
  Astronomical Site Evaluation in the Visible and Radio Range,
  ed.  J. Vernin, Z. Benkhaldoun, \& C. M\~{u}noz-Tu\~{n}on, 
  (San Francisco: Astr. Soc. of the Pacific)

\bibitem[{Chamberlin \& Bally 1994}]{chamberlin94}
Chamberlin, R.~A., \& Bally, J. 1994, Applied Optics, 33(6), 1095

\bibitem[{Chamberlin \& Bally 1995}]{chamberlin95}
Chamberlin, R.~A., \& Bally, J. 1995, Int. J. Infrared and Millimeter Waves, 16,
  907

\bibitem[Chamberlin {et~al.} 1997]{chamberlin97}
Chamberlin, R.~A., Lane, A.~P., \& Stark, A.~A. 1997, \apj, 476, 428

\bibitem[{Engargiola {et~al.} 1994}]{engargiola94}
Engargiola, G., Zmuidzinas, J., \& Lo, K.-Y. 1994, Rev. Sci. Instr., 65, 1833

\bibitem[{Gerecht {et~al.} 1999}]{gerecht99}
Gerecht, E., Musante, C.~F., Zhuang, Y., Yngvesson, K.~S., Goyette, T.,
  Dickinson, J., Waldman, J., Yagoubov, P.~A., Gol'tsman, G.~N., Voronov,
  B.~M., \& Gershenzon, E.~M. 1999, IEEE Trans., MTT-47, 2519

\bibitem[{Groppi {et~al.} 2000}]{groppi00}
Groppi, C., Walker, C., Hungerford, A., Kulesa, C., Jacobs, K., \& Kooi, J.
  2000, in ASP Conference Series, Vol 217, Imaging at Radio Through Submillimeter Wavelengths, ed. J.~G. Mangum
  \& S.~J.~E. Radford, (San Francisco: Astr. Soc. of the Pacific), 48


\bibitem[{Honingh {et~al.} 1997}]{honingh97}
Honingh, C.~E., Hass, S., Hottgenroth, K., Jacobs, J., \& Stutzki, J. 1997,
  IEEE Trans. Appl. Superconductivity, 7, 2582

\bibitem[{Keene {et~al.} 1998}]{keene98}
Keene, J., Schilke, P., Kooi, J., Lis, D.~C., Mehringer, D.~M., \& Phillips,
  T.~G. 1998, \apj, 494, L107

\bibitem[{Kim {et~al.} 2000}]{kim00}
Kim, S., Martin, C.~L., Stark, A.~A., \& Lane, A.~P. 2000, American
  Astronomical Society Meeting 197, BAAS, 32, 4.04

\bibitem[{Kooi {et~al.} 1995}]{kooi95}
Kooi, J.~W., Chan, M.~S., Bumble, B., LeDuc, H.~G., Schaffer, P.~L., \&
  Phillips, T.~G. 1995, Int. J. IR and MM Waves, 16

\bibitem[{Kooi {et~al.} 1992}]{kooi92}
Kooi, J.~W., Man, C., Phillips, T.~G., Bumble, B., \& LeDuc, H.~G. 1992, IEEE
  Trans. Microwaves Theory and Techniques, 40, 812

\bibitem[{Lane 1998}]{lane98}
Lane, A.~P. 1998, in ASP Conference Series, Vol. 141, Astrophysics from Antarctica,
  ed. G.~Novak \& R.~H. Landsberg, (San Francisco: Astr. Soc. of the Pacific), 289

\bibitem[{{Lay} \& {Halverson} (2000)}]{lay99}
{Lay}, O.~P., \& {Halverson}, N.~W. 2000, \apj, 543, 787

\bibitem[{NRC 2001}]{nrc01}
National Research Council  2001, Astronomy and Astrophysics in the New
  Millennium, (National Academy Press)

\bibitem[{Pardo {et~al.} 2001}]{pardo01b}
Pardo, J.~R., Cernicharo, J., \& Serabyn, E. 2001, IEEE Trans. Antennas and
  Propagation, 49, 1683

\bibitem[{Schieder {et~al.} 1989}]{Schieder89}
Schieder, R., Tolls, V., \& Winnewisser, G. 1989, Exp. Astron., 1, 101

\bibitem[{Schwerdtfeger (1984)}]{schwerdtfeger}
Schwerdtfeger, W. 1984, Weather and Climate of the Antarctic, (Amsterdam:
  Elsevier)

\bibitem[{Stark {et~al.} (2001)}]{stark01}
Stark, A.~A., Bally, J., Balm, S.~P., Bania, T.~M., Bolatto, A.~D., Chamberlin,
  R.~A., Engargiola, G., Huang, M., Ingalls, J.~G., Jacobs, K., Jackson, J.~M.,
  Kooi, J.~W., Lane, A.~P., Lo, K.-Y., Marks, R.~D., Martin, C.~L., Mumma, D.,
  Ojha, R., Schieder, R., Staguhn, J., Stutzki, J., Walker, C.~K., Wilson,
  R.~W., Wright, G.~A., Zhang, X., Zimmermann, P., \& Zimmermann, R. 2001,
  \pasp, 113, 567

\bibitem[{Stark {et~al.} (1997) }]{stark97a}
Stark, A.~A., Chamberlin, R.~A., Cheng, J., Ingalls, J., \& Wright, G. 1997,
  Rev. Sci. Instr., 68, 2200

\bibitem[{Swain {et~al.} 1998}]{swain98}
Swain, M.~R., Bradford, C.~M., Stacey, G.~J., Bolatto, A.~D., Jackson, J.~M.,
  Savage, M., \& Davidson, J.~A. 1998, SPIE, 3354, 480

\bibitem[{Tieftrunk {et~al.} 2001}]{tieftrunk01}
Tieftrunk, A.~R., Jacobs, K., Martin, C.~L., Siebertz, O., Stark, A.~A.,
  Walker, C.~K., \& Wright, G.~A. 2001, \aap, 375, L23


\bibitem[{Walker {et~al.} 1992}]{walker92}
Walker, C.~K., Kooi, J.~W., Chan, W., LeDuc, H.~G., Schaffer, P.~L., Carlstrom,
  J.~E., \& Phillips, T.~G. 1992, Int. J. Infrared and Millimeter Waves, 13,
  785


\bibitem[{Yngvesson {et~al.} 2001}]{yngvesson01}
Yngvesson, K.~S., Musante, C.~F., Ji, M., Rodriguez, F., Zhuang, Y., Gerecht,
  E., Coulombe, M., Dickinson, J., Goyette, T., Waldman, J., Walker, C.~K.,
  Stark, A.~A., \& Lane, A.~P. 2001, in Twelfth Intern. Symp. Space THz
  Technology, (I. Medhi, Ed.), San Diego, CA, Feb. 14-16, 2001, p. 262


\bibitem[{Zmuidzinas \& LeDuc 1992}]{zmuidzinas92}
Zmuidzinas, J., \& LeDuc, H.~G. 1992, IEEE Trans. Microwave Theory Tech., 40,
  1797

\end{thebibliography}
\end{document}